# OMP2HMPP: Compiler Framework for Energy-Performance Trade-off Analysis of Automatically Generated Codes

Albert Saà-Garriga[1], David Castells-Rufas[2] and Jordi Carrabina[3]

[1] Dept. Microelectronics and Electronic Systems, Universitat Autònoma de Barcelona
Bellaterra, Barcelona 08193, Spain.

**Abstract**
We present OMP2HMPP, a tool that, in a first step, automatically translates OpenMP code into various possible transformations of HMPP. In a second step OMP2HMPP executes all variants to obtain the performance and power consumption of each transformation. The resulting trade-off can be used to choose the more convenient version.
After running the tool on a set of codes from the Polybench benchmark we show that the best automatic transformation is equivalent to a manual one done by an expert. Compared with original OpenMP code running in 2 quad-core processors we obtain an average speed-up of 31× and 5.86× factor in operations per watt.

## 1. Introduction

High-performance computers are based more and more in heterogeneous architectures. The specialization of computational units and the adoption of different models of computation in various nodes, allow systems to increase performance and energy efficiency. GPGPUs were discovered as promising vehicles for general-purpose high-performance computing and have become a popular part of such heterogeneous architectures since these give in most of the studied cases a great speed-up compared to parallel CPU versions. Accelerators (Nvidia GPU, Intel Xeon Phi...) are gaining market share: 28% of systems have one or more accelerators installed (2011/2012 survey [16]). There is a continuous increase in the number of topics that are interested in accelerators as a way to compute their calculations faster. However, the effort needed to program them might become a hurdle for their wider adoption. Some languages have been to offer programmability for general purpose computing i.e. Compute Unified Device Architecture (CUDA) [20], HMPP [[17], [5]], RapidMind [15], PeakStream [22] or, CTM [19]. However, GPGPUs programming alternatives are complex and error-prone, compared to programming general purpose CPUs and parallel programming models such as OpenMP [22]. In this paper we present automatic transformation tool on the source code oriented to work with General Propose Units for Graphic Processing(GPGPUs). This new tool (OMP2HMPP) is in the category of source-to-source code transformations and seek to facilitate the generation of code for GPGPUs. OMP2HMPP is able to run specific sections on accelerators, so that the program executes more efficiently on heterogeneous platform. OMP2HMPP was initially developed for the systematic exploration of the large set of possible transformations to determine the optimal one in terms performance (both energy and time consumption). OMP2HMPP grows upon with knowledge of the existing alternatives for GPGPUs programming, studying the use of the most promising for its output (HMPP). Since HMPP offers the easiest way to apply the migration because is a directive-based language. Meta-information added in the source code of the application does not change the semantic of the original code thus simplifying the kernel creation. Additionally it offers an incremental way of migrating applications by first declaring and generating kernels of critical to later manage data transfers and finishing by optimizing kernel performance and data synchronization. They address the remote execution (RPC) of functions or regions of code on GPUs and many-core accelerators as well as the transfer of data to and from the target device memory. In addition, most of the other alternatives rely on a stream programming style but a program written for a given platform cannot run on another one. HMPP takes a radically different approach. A HMPP application can be compiled with an off-the-shelf compiler and run without any specific run-time to produce a conventional native binary. Moreover, thanks to its dynamic linking mechanism, a HMPP application is able to make use of either a new accelerator or an improved codelet without having to recompile the application source. This way we aim at preserving legacy codes and insulate them from frequent hardware platform changes that tend to characterize hybrid multi-cores, e.g. fast GPU architecture evolution. As we aforementioned, the existing programming alternatives(HMPP inclusive) for GPGPUs programming, are still more complex, due to the hardware complexity, than programming general-purpose CPUs and parallel programming models such as OpenMP. For this reason OMP2HMPP is thought to avoid this task to the programmer. To simplify the transformation task, OMP2HMPP will reuse existing source codes that were designed to describe parallelism using OpenMP directives. High-performance computing (HPC) community which has been commonly using a couple of standards: MPI and OpenMP. OpenMP is better oriented to support of multi-platform shared memory parallel programming in C/C++. It defines an interface for parallel applications on wide platforms range: form the desktop to the supercomputer. As a natural extension, OpenMP could be combined with the use of HMPP to study the performance and energy efficiency trade-off.

On those bases, we developed OMP2HMPP tool which is able to:

—Run specific sections of the proposed code on accelerators. OMP2HMPP combines the use of CPU parallel code and GPU so that the program can execute more efficiently on a heterogeneous platform.

—Do systematic exploration of the large set of possible transformations from pragma-annotated code to determine the optimal transformation.

—Introduce GPGPU acceleration that could not be optimal but would provide a good trade-off between performance and development effort, avoiding going through new learning curve. It will give a comparison of all the possible combinations of OpenMP and HMPP configurations, in terms of time and energy spent executing each code section.

OMP2HMPP is a Source to Source compiler (S2S) based on BSCs Mercurium framework [18] that generates HMPP code from OpenMP. Mercurium [3] gives us a source-to-source compilation infrastructure aimed at fast prototyping and supports C and C++ languages. This platform is mainly used in the Nanos environment to implement OpenMP but since it is quite extensible it has been used to implement other programming models or compiler transformations.

Extending Mercurium is achieved using a plugin architecture, where plugins represent several phases of the compiler. These plugins are written in C++ and dynamically loaded by the compiler according to the selected configuration. Code transformations are implemented to the source code (there is no need to know or modify the internal syntactic representation of the compiler).

OMP2HMPP uses Mercurium to implement our S2S transformation phases, providing OMP2HMPP with an abstract representation of the input source code: the Abstract Syntax Tree (AST). AST provides an easy access to source code structure representation, the table of symbols and the context of these.

### 1.1 HMPP Directives

The proposed tool is able to combine the use of the following HMPP directives with the original OpenMP directives:

—**Callsite**: Specifies the use of a codelet at a given point in the program. Related data transfers and synchronization points that are inserted elsewhere in the application have to use the same label.

—**Codelet**: Specifies that a version of the function following must be optimized for a given hardware.

—**Group:** Allows the declaration of a group of codelet.

—**Advanced Load:** Uploads data before the execution of the codelet.

—**Delegate Store:** Represents the opposite of the advancedload directive in the sense that it downloads output data from the HWA to the host.

—**Synchronize:** Specifies to wait until the completion of an asynchronous callsite execution.

—**Release:** Specifies when to release the HWA for a group or a stand-alone codelet.

—**No Update:** This property specifies that the data is already available on the HWA and so that no transfer is needed. When this property is set, no transfer is done on the considered argument.

—**Target:** Specifies one or more targets for which the codelet must be generated. It means that according to the target specified, if the corresponding hardware is available AND the codelet implementation for this hardware is also available, this one will be executed. Otherwise, the next target specified in the list will be tried. OMP2HMPP always use CUDA since we will test it in a server without OpenCL support.

With these directives OMP2HMPP is able to create a version that in the most of the cases rarely will differs from a hand-coded HMPP version of the original problem.

### 1.2 Related Work

Dominant GPU programming models have traditionally been CUDA and OpenCL [8] Working Group 2008. In recent years; many source-to-source compiler alternatives have been proposed to overcome the GPGPU programming complexity. Among them, some that are similar to the tool proposed in this paper are presented for discussion. Opposite to OMP2HMPP, the following methods produce direct transformation to CUDA language, not to HMPP, which means that the CUDA programming complexity is directly exposed to the final user.

Some of the proposals extend, in one way or another, the current standards such a C/C++, OpenMP, etc. 2014]. On the other hand, there are proposals that do not require any language extension to transform the source code directly from CPU to GPUs.

One of the examples that include language extensions is proposed in [21]. CAPS, CRAY, Nvidia and PGI, (members of the OpenMP Language Committee) published OpenACC in November 2011. OpenACC has been proposed as standard for directive-based standard programming as it contributes to the specification of OpenMP for accelerators. In the same way, but not with the same consensus that OpenACC, in [10], a programming interface called OpenMPC is presented. This paper shows and extensive analysis of the actual state of the art in OpenMP to CUDA source-to-source compilers and CUDA optimizer. OpenMPC provides an abstraction of the complexity of CUDA programming model and increases its automation though user-assistance tuning system. Both, OpenMPC and OpenACC, require time to understand the new proposed directives, and to manually optimize the data transfer between CPU and GPU. Opposite to both, OMP2HMPP adds just two new OpenMP directives and the programmer can forgets to deal with new languages and their underlying optimization. Another option is hiked directive-based language [6], which is a set of directives for CUDA computation and data attributes in a sequential program. However, hiked has the same programming paradigm than CUDA; even though it hides the CUDA language syntax, the complexity of the CUDA programming and memory model is directly exposed to programmers. Moreover, in contrast to OMP2HMPP, hiked does not provide any transfer optimization. Finally [11] and [1], propose an OpenMP compiler for hybrid CPU/GPU computing architecture. In these papers they propose to add a directive to OpenMP in order to choose where the OpenMP block must be executed (CPU/GPU). The process is full hide to the programmer and is a direct translated to CUDA. Again, it does not provide any transfer optimization. There are fewer proposals that try to directly transform C/C++ code to CUDA without the need of any new language extension. [4] present a tool that uses unimodular loop transformation theory to analyze the loops that could be transformed to work in parallel kernels either OpenMP or CUDA trough ROSE[14] compiler. Par4All [1] transform codes originally wrote in C or FORTRAN to OpenMP, CUDA or OpenCL. Par4All uses polyhedral model for analysis and transforms C/C++ source code and adds OpenMP directives where the program thinks that can be useful. This transformation allows the re-factorization of the newly created OpenMP blocks to GPGPUs kernels by moving OpenMP directives to CUDA language. However, this transformation does not take into account the kernel data-flow context and this lead to non-optimal results in data-transfers. Nevertheless, both tools tool, and the tools studied in [7] could be useful to transform sequential codes to be input codes in OMP2HMPP tool.

For source-to-source compiler infrastructure, there are many possible solutions as LLVM [9], PIPS , Cetus, ROSE and the used Mercurium since it gives us support to C/C++ source codes is more friendly to work with his intermediate representation trough an well documented API that allows extensions in that one as was demonstrated in [12] and [13].

## 2. S2S Transformations

OMP2HMPP implementation is a pipeline of S2S compiler transformations that transform OpenMP code to HMPP. This pipeline has two compiler phases to devote to following transformation steps (Outline and Inline).

The first phase (outline phase) transforms pragma OpenMP block into HMPP codelet and callsite. In order to solve problems that can appear related to the use of global variables inside HMPP kernels and with the call of functions not allowed in the device, the second compiler phase will check the scope of all the variables used in codelet and, at the same time will transform function calls doing an inline inside the codelet (inline phase). Figure 1 shows the work-flow of the transformation process. This procedure is detailed in the following subsections.

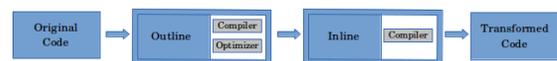

Figure 1: S2S Transformation Phases

Table 1: S2S Transformation Process

| OPENMP | HMPP |
|---|---|
| ```
int main()
{
...
#pragma omp parallel for check
 for(i=0;i<row;++i){
  for(j=0);j<col;++j){
   result[i][j] = 0;
   array[i*j] = mat[i][j];
   for(k=0;k<row;++k){
    a=0;
    while(a<10) {
     result[i][j] += mat1[i][k]*mat2[k][j]*array[i*j];
     a++
    }
   }
  }
 }
...
}
``` | ```
#pragma hmpp _intr_for__ol_3_main codelet, target = CUDA,
            args[result,array].io=inout, args[array].size={row*col}, &
#pragma hmpp & args[*].transfer=auto
void _intr_for__ol_3_main(int i, int row, int j, int result[row][col],
    int *array, int mat1[row][col], int k, int a, int mat2[row][col])
{
for(i=0;i<row;++i){
  for(j=0);j<col;++j){
   result[i][j] = 0;
   array[i*j] = mat[i][j];
   for(k=0;k<row;++k){
    a=0;
    while(a<10) {
     result[i][j] += mat1[i][k]*mat2[k][j]*array[i*j];
     a++
}}}}}
int main() {
...
#pragma hmpp _inst_for__ol_3_main callsite
_instr_for__ol_3_main(i,row,j,col,result,array,mat1,k,a,mat2);
...
}
``` |

OMP2HMPP generates multiple implementations which differ in the use of different HMPP pragma parameter configurations, then compiles and executes them collecting the elapsed time and energy consumed in every execution. Results can be plotted in order to obtain the trade-off curves that will allow obtaining the optimal working point. These processes performed after source to source transformation process are described in Figure 2.

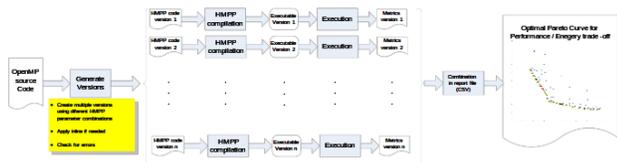

Figure 2: S2S Transformation Process

## 2.1 Outline Phase

The outline phase is responsible of the inline transformation (transforming OpenMP blocks to HMPP kernels and calls). Such phase find all the OpenMP instructions in the input code, and then detect the start and end of the pragma block.

Once this detection is finished, the outline phase declares a function with the same functionality of the pragma block. Table 1 gives an example of the transformation implemented in this phase. The left side shows the original code, while the right side shows the code after outlining transformations.

The outline phase is divided in two stages: The first one is named compilation stage, and is liable for the source-to-source transformation, the understand of the programmer source code and, the generation of different versions. The second stage is called optimization, and is devoted to improvement of the code proposed by the programmer.

### 2.1.1 Compilation

The compiler has to deal with all the OpenMP standard directives i.e. shared, private, reduction, and so on. There are two possible scenarios, OpenMP blocks can be expressed as simple blocks where OMP2HMPP tool will transform this blocks into HMPP codelet, or a group of blocks where the compilation will transform OpenMP parallel blocks either into HMPP codelet groups, and therefore share variables between these codelet groups, or will divide in simple blocks being able to specify individually where these blocks will be computed as is shown in Table 2, exploring the capabilities of the heterogeneous architecture.

Each of the found OpenMP blocks will be transformed generating a new source code version, and moreover OMP2HMPP explore all the possible HMPP configurations that can be used in these blocks, that implies that the number of generated versions will grow exponentially as we have more OpenMP blocks to transform. To solve that, we create two new OpenMP directives that give to the user the possibility to generate the final chosen version in doing smaller explorations of all the possible versions that can be generated. These directives allow the user to explore the generated versions block by block. The set of new directives are described in the following list and exemplified in Figure 3.





Table 2: OpenMP Block Division

| OPENMP | HMPP |
|---|---|
| ```
int main()
{
    ...
    #pragma omp parallel shared(myTableOut,myTable) check
    for (; (index < iterations); index++ )
    {
        #pragma omp for
        for (i=SPANI; i < WORKSIZE - SPANI; i++) {
            for (j=SPANJ; j < LINESIZE - SPANJ; j++) {
                ...
            }
        }
        theDiffNorm = 0.0;
        diffsum=theDiffNorm;
        #pragma omp for reduction(+:diffsum)
        for (i = 1;i < (1 + MAXM + 1) - 1; i++) {
            for (j = 1;j < (1 + MAXN + 1) - 1;j++) {
                …
            }
        }
        theDiffNorm=diffsum;
    }
    #pragma omp parallel for reduction(+:diffsum) fixed(10,1,0)
    for (i = 1;i < (1 + MAXM + 1) - 1; i++){
        for (j = 1;j < (1 + MAXN + 1) - 1;j++){
            ...
        }
    }
    displayRegion( myTable);
    return 0;
}
``` | ```
int main()
{
    #pragma hmpp <group1> group, target=CUDA
    #pragma hmpp <group1> mapbyname, myTableOut
    ...
    #pragma hmpp <group1> _instr_for4_ol_13_main advancedload,
            args[myTable], args[myTableOut].addr="myTableOut"
    int a = 0;
    double diffsum = 0.0;
    for (; (index < iterations);index++)  {
    #pragma hmpp <group1> _instr_for4_ol_13_main callsite,
                args[myTableOut].noupdate=true
        _instr_for4_ol_13_main(i, j, a, myTable, myTableOut);
        theDiffNorm = 0.0;
        diffsum = theDiffNorm;
    #pragma hmpp <group1> _instr_for4_ol_13_main delegatedstore,
                args[myTableOut],
    args[myTableOut].addr="myTableOut"
    #pragma omp parallel for reduction(+:diffsum) shared(myTable)
    private(i, j)
    for (i = 1; i < (1 + 5000 + 1) - 1; i++) {
        for (j = 1; j < (1 + 5000 + 1) - 1; j++) {
            ...
        }
    }
    theDiffNorm = diffsum;
    }
    #pragma hmpp <group1> _instr_for4_ol_20_main callsite,
                args[myTableOut].noupdate=true
    _instr_for4_ol_20_main(i, j, myTableOut, myTable, a, &diffsum);
    displayRegion(myTable);
    #pragma hmpp <group1> release
    return 0;
}
``` |

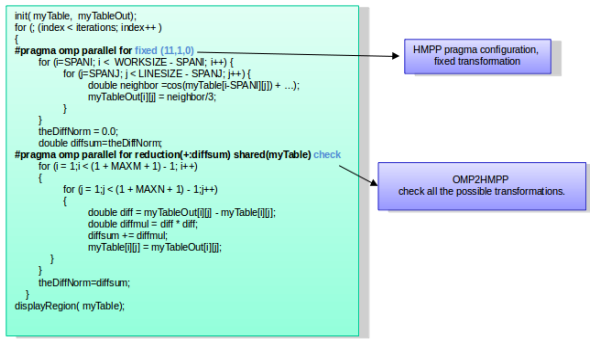

Figure 3: New Directives Explanation

—**CHECK**: The OpenMP blocks with a pragma set #pragma omp parallel for check, will be transformed using all the HMPP possible configurations, including the no-transformation, which will keep the actual OpenMP block as in the original code.

—**FIXED**: The programmer will fix a transformation for a certain block by giving information flags for the next execution of OMP2HMPP. I.e. *#pragma omp parallel for fixed (10, 1, 0)*.

The programmer can explore all the possible configurations of HMPP for a certain OpenMP block, and with the report generated by OMP2HMPP decide which is the best configuration, as will be explained in section 3. This configuration could be established for the following executions of OMP2HMPP using the FIXED directive. FIXED directive is full complemented with a set of three flags that are described in Figure 4. This flags have an internal binary representation that we transform into decimal one in order to compress the length in OpenMP pragma instruction specification.

Table 3: OpenMP Reduction Directive

| OPENMP | HMPP |
|---|---|
| ```
#pragma omp parallel for reduction(+:diffsum) shared(myTable) check
for (i = 1;i < (1 + MAXM + 1) - 1; i++)
{
    for (j = 1;j < (1 + MAXN + 1) - 1;j++)
    {
        double diff = myTableOut[i][j] - myTable[i][j];
        double diffmul = diff * diff;
        diffsum += diffmul;
        myTable[i][j] = myTableOut[i][j];
    }
}
``` | ```
#pragma hmpp _instr_for__ol_75_main codelet, target = CUDA,
            args[myTable].io=inout, args[myTableOut].io=in,
            args[diffsum_reduced].io=inout,
args[diffsum_reduced].size={1}
void _instr_for__ol_75_main(int i, int j, double
myTableOut[5002][5002], double myTable[5002][5002],
                            double *diffsum_reduced)
{
    double diffsum = *diffsum_reduced;
    #pragma hmppcg gridify(i, j), reduce(+:diffsum)
    for (i = 1; i < (1 + 5000 + 1) - 1; i++)
    {
        for (j = 1; j < (1 + 5000 + 1) - 1; j++)
        {
            double diff = myTableOut[i][j] - myTable[i][j];
            double diffmul = diff * diff;
            diffsum += diffmul;
            myTable[i][j] = myTableOut[i][j];
        }
    }
    *diffsum_reduced = diffsum;
}
``` |





Table 4: Example of Contextual Analysis with noUpdate directive codelet

```
HMPP

#pragma hmpp <group0_12> _instr_for12_ol_12_main codelet, args[myTable, myTableOut].io=in
void _instr_for42_ol_12_main(int i, int j, double myTable[5002][5002], double myTableOut[5002][5002])
{
#pragma hmppcg gridify(i, j)
    for (i = 1;
         i < (1 + 5000 + 1) - 1;
         i++)
    {
        for (j = 1;
             j < (1 + 5000 + 1) - 1;
             j++)
        {
            double neighbor = cos(myTable[i - 1][j]) + sin(myTable[i][j - 1]) + sin(myTable[i][j + 1]) + cos(myTable[i + 1][j]);
            myTableOut[i][j] = neighbor / 3;
        }
    }
}
```

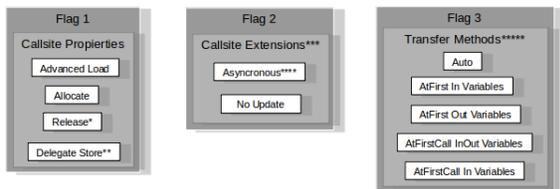

(*) Just available in case that a group of codelets is created.
(**) Implies release flag 1 active.
(***) Just can be sets when Advanced load flag is active.
(****) Implies Delegate store flag 1 active.
(*****) Just applied when flag 1 and 2 are sets in 0 in simple callsite.

Figure 4: FIXED Directive Flags Explanation

The detection for each variable being passed as value, copy or reference (see example in III) allows distinguishing the context of the variables inside every block. This allows the exploration of all the possible transformations for one OpenMP block OMP2HMPP.

Also, OMP2HMPP has to treat in different ways array and matrix parameter passing .HMPP distinction requirements impose detecting when both of them are used, as shown in me.

The compilation stage take more control of how the outlived kernel is divided detecting the variables that define the two outer for loops and, using this information, define the grid division with the use of the hmppcg gridify directive. We illustrate an example of this case in Table 3.

As aforementioned, OMP2HMPP deals with all the possible OpenMP directives, to show an example of this we shown how it works with the use of the reduction directive, which is a safe way of joining work from all threads after construct. OMP2HMPP simulate the pass by reference with variable in reduction directive diffsum. The result of this transformation is shown in Table 3.

Table 5: Example of Contextual Analysis with noUpdate directive calls.

| OPENMP | HMPP |
|---|---|
| ```
int main()
{
  int index =0;
  double theDiffNorm = 1;
  double RefDiffNorm = 0;
  int iterations = 99;
  int worksize=WORKSIZE, linesize=LINESIZE;
  int i,j,o,a;
  double diffsum,diff,diffmul;
  init( myTable, myTableOut);
  for (index=0; (index < iterations); index++ )
  {
#pragma omp parallel for shared(myTableOut) check
    for (i=SPANI; i < WORKSIZE - SPANI; i++) {
      for (j=SPANJ; j < LINESIZE - SPANJ; j++) {
        double neighbor=cos(myTable[i-SPANI][j]) +sin(myTable[i][j-SPANJ])
                       +sin(myTable[i][j+SPANJ]) +cos(myTable[i+SPANI][j]);
        myTableOut[i][j] = neighbor/3;
      }
    }
    theDiffNorm = 0.0;
    diffsum=theDiffNorm;
#pragma omp parallel for reduction(+:diffsum) shared(myTable) check
    for (i = 1;i < (1 + MAXM + 1) - 1; i++)
    {
      for (j = 1;j < (1 + MAXN + 1) - 1;j++)
      {
        diff = myTableOut[i][j] - myTable[i][j];
        diffmul = diff * diff;
        diffsum += diffmul;
        myTable[i][j] = myTableOut[i][j];
      }
    }
    theDiffNorm=diffsum;
  }
  displayRegion( myTable);
  return 0;
}
``` | ```
int main()
{
#pragma hmpp <group0_12> group, target=CUDA
#pragma hmpp <group0_12> mapbyname, myTable,myTableOut
  int index = 0;
  double theDiffNorm = 1;
  double RefDiffNorm = 0;
  int iterations = 99;
  int worksize = (1 + 5000 + 1), linesize = (1 + 5000 + 1);
  int i, j, o, a;
  double diffsum, diff, diffmul;
  init(myTable, myTableOut);
#pragma hmpp <group0_12> _instr_for12_ol_12_main advancedload,
args[myTable, myTableOut], args[myTable].addr="myTable",
                                       args[myTableOut].addr="myTableOut"
  for (index = 0;
       (index < iterations);
       index++)
  {
#pragma hmpp <group0_12> _instr_for12_ol_12_main callsite,
                        args[myTable, myTableOut].noupdate=true
    _instr_for12_ol_12_main(i, j, myTable, myTableOut);
    theDiffNorm = 0.0;
    diffsum = theDiffNorm;
#pragma hmpp <group0_12> _instr_for12_ol_17_main callsite,
                        args[myTableOut, myTable].noupdate=true
    _instr_for12_ol_17_main(i, j, diff, myTableOut, myTable,
diffmul, &diffsum, a);
    theDiffNorm = diffsum;
  }
#pragma hmpp <group0_12> _instr_for12_ol_17_main delegatedstore,
                        args[myTable], args[myTable].addr="myTable"
  displayRegion(myTable);
#pragma hmpp <group0_12> release
  printf("theDiffNorm:%.12g RefDiffNorm=%.12g;", theDiffNorm,
RefDiffNorm);
  return 0;
}
``` |

2.1.2 Optimization

This stage improves the code proposed by the programmer by exploiting some context situations. The optimization reduces the number of transfers between CPU and GPU understanding the variable context of the OpenMP kernels.

In order to implement the optimization task, OMP2HMPP must perform an accurate contextual analysis of the original code. For that, OMP2HMPP does a contextual information study, taking care of each array or matrix variables needed in each of the marked OpenMP kernels that have to be transformed. Through this analysis OMP2HMPP is able to understand the following variable context:

—determine the kind access (write/read)
—Determine the host where is used (CPU/GPU)
—The scope of the instruction where the variable is used (Loop detection)

We show a simple example of that context understanding in Figure 5. Variables A and C are used inside the OpenMP block. OMP2HMPP uses that information to select the best use of HMPP directives that minimize the number of data transfers. In Figure 5, OMP2HMPP has two variables to analyze A and C. In the case of A, A has to be uploaded to GPU, but is not necessary to download it after the kernel call because there is no read of that variable until the end of the code. In the case of variable C, C has to be downloaded from GPU to CPU, but there is no need to upload that to GPU since the kernel do not do a read of C inside. With that information, OMP2HMPP will use an advancedload in the case of A and will put that directive as close as possible to the last rite expression, to optimize the data transfer and improve the performance of the generated code as shown in Figure 6a. In the case of C, OMP2HMPP will put a delegatestore directive, as far as possible of the kernel call, and that will increase the performance of the generated code, as is shown in Figure 8a. Figure 6b and Figure 8b illustrate the use of a bad transfer policy in the same problems.

Table 6: Loop Dealing Example

| OPENMP | HMPP |
|---|---|
| ```c
int main()
{
    int index =0;
    double theDiffNorm = 1;
    double RefDiffNorm = 0;
    int iterations = 99;
    int worksize=WORKSIZE, linesize=LINESIZE;
    int i,j,o,a;
    double diffsum,diff,diffmul;
    init( myTable, myTableOut);
    for (index=0; (index < iterations); index++ )
    {
    #pragma omp parallel for shared(myTableOut)
        for (i=SPANI; i < WORKSIZE - SPANI; i++) {
            for (j=SPANJ; j < LINESIZE - SPANJ; j++) {
                double neighbor=cos(myTable[i-SPANI][j]) +sin(myTable[i][j-SPANJ])
                    +sin(myTable[i][j+SPANJ]) +cos(myTable[i+SPANI][j]);
                myTableOut[i][j] = neighbor/3;
            }
        }
        theDiffNorm = 0.0;
        diffsum=theDiffNorm;
        a=0;
    #pragma omp parallel for reduction(+:diffsum) shared(myTable) check
        for (i = 1;i < (1 + MAXM + 1) - 1; i++)
        {
            for (j = 1;j < (1 + MAXN + 1) - 1;j++)
            {
                diff = myTableOut[i][j] - myTable[i][j];
                diffmul = diff * diff;
                diffsum += diffmul;
                a=2;
                myTable[i][j] = myTableOut[i][j];
            }
        }
        theDiffNorm=diffsum;
    }
    displayRegion( myTable);
    return 0;
}
``` | ```c
int main()
{
#pragma hmpp <group0_46> group, target=CUDA
#pragma hmpp <group0_46> mapbyname, myTable
    int index = 0;
    double theDiffNorm = 1;
    double RefDiffNorm = 0;
    int iterations = 99;
    int worksize = (1 + 5000 + 1), linesize = (1 + 5000 + 1);
    int i, j, o, a;
    double diffsum, diff, diffmul;
    for(int l=0;l<20;l++) {
        init(myTable, myTableOut);
    }
#pragma hmpp <group0_46> _instr_for46_ol_17_main advancedload, args[myTable], args[myTable].addr="myTable"
    for (index = 0;(index < iterations); index++)
    {
#pragma omp parallel for shared(myTableOut) private(i,j)
        for (i = 1; i < (1 + 5000 + 1) - 1; i++)
        {
            for (j = 1; j < (1 + 5000 + 1) - 1; j++)
            {
                double neighbor = cos(myTable[i - 1][j])
                    + sin(myTable[i][j - 1])
                    + sin(myTable[i][j + 1])
                    + cos(myTable[i + 1][j]);
                myTableOut[i][j] = neighbor / 3;
            }
        }
        theDiffNorm = 0.0;
        diffsum = theDiffNorm;
        a = 0;
#pragma hmpp <group0_46> _instr_for46_ol_17_main callsite, args[myTable].noupdate=true, asynchronous
        _instr_for46_ol_17_main(i, j, diff, myTableOut, myTable, diffmul, &diffsum, a);
#pragma hmpp <group0_46> _instr_for46_ol_17_main synchronize
#pragma hmpp <group0_46> _instr_for46_ol_17_main delegatestore, args[diffsum_reduced],
args[diffsum_reduced].addr="&diffsum"
#pragma hmpp <group0_46> _instr_for46_ol_17_main delegatestore, args[myTable],
args[myTable].addr="myTable"
        theDiffNorm = diffsum;
    }
    displayRegion(myTable);
#pragma hmpp <group0_46> release
    printf("theDiffNorm:%.12g RefDiffNorm=%.12g;", theDiffNorm, RefDiffNorm);
    return 0;
}
``` |





Table 7: Example of Contextual Analysis with noUpdate directive calls.

| OPENMP | HMPP |
|---|---|
| ```
int main()
{
  int index =0;
  double theDiffNorm = 1;
  double RefDiffNorm = 0;
  int iterations = 99;
  int worksize=WORKSIZE, linesize=LINESIZE;
  int i,j,o,a;
  double diffsum,diff,diffmul;
  init( myTable,  myTableOut);
  for (index=0; (index < iterations); index++ )
  {
    #pragma omp parallel for shared(myTableOut) check
      for (i=SPANI; i <  WORKSIZE - SPANI; i++) {
        for (j=SPANJ; j < LINESIZE - SPANJ; j++) {
          double   neighbor=cos(myTable[i-SPANI][j])  +sin(myTable[i][j-SPANJ])
                           +sin(myTable[i][j+SPANJ])
  +cos(myTable[i+SPANI][j]);
          myTableOut[i][j] = neighbor/3;
        }
      }
    theDiffNorm = 0.0;
    diffsum=theDiffNorm;
    #pragma omp parallel for reduction(+:diffsum) shared(myTable) check
      for (i = 1;i < (1 + MAXM + 1) - 1; i++)
      {
        for (j = 1;j < (1 + MAXN + 1) - 1;j++)
        {
          diff = myTableOut[i][j] - myTable[i][j];
          diffmul = diff * diff;
          diffsum += diffmul;
          myTable[i][j] = myTableOut[i][j];
        }
      }
    theDiffNorm=diffsum;
  }
  displayRegion( myTable);
  return 0;
}
``` | ```
int main()
{
#pragma hmpp <group0_12> group, target=CUDA
#pragma hmpp <group0_12> mapbyname, myTable,myTableOut
    int index = 0;
    double theDiffNorm = 1;
    double RefDiffNorm = 0;
    int iterations = 99;
    int worksize = (1 + 5000 + 1), linesize = (1 + 5000 + 1);
    int i, j, o, a;
    double diffsum, diff, diffmul;
    init(myTable, myTableOut);
#pragma   hmpp  <group0_12>  _instr_for12_ol_12_main    advancedload,
args[myTable,myTableOut], args[myTable].addr="myTable",
                                       args[myTableOut].addr="myTableOut"
    for (index = 0;
        (index < iterations);
        index++)
    {
#pragma hmpp <group0_12> _instr_for12_ol_12_main callsite,
                 args[myTable, myTableOut].noupdate=true
      _instr_for12_ol_12_main(i, j, myTable, myTableOut);
      theDiffNorm = 0.0;
      diffsum = theDiffNorm;
#pragma hmpp <group0_12> _instr_for12_ol_17_main callsite,
                 args[myTableOut, myTable].noupdate=true
      _instr_for12_ol_17_main(i,   j,  diff,   myTableOut,   myTable,
diffmul,  &diffsum,  a);
      theDiffNorm = diffsum;
    }
#pragma hmpp <group0_12> _instr_for12_ol_17_main delegatedstore,
                 args[myTable], args[myTable].addr="myTable"
    displayRegion(myTable);
#pragma hmpp <group0_12> release
    printf("theDiffNorm:%.12g     RefDiffNorm=%.12g;",     theDiffNorm,
RefDiffNorm);
    return 0;
}
``` |

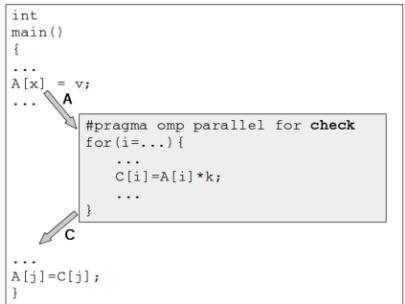

Figure 5: Context Analysis Example

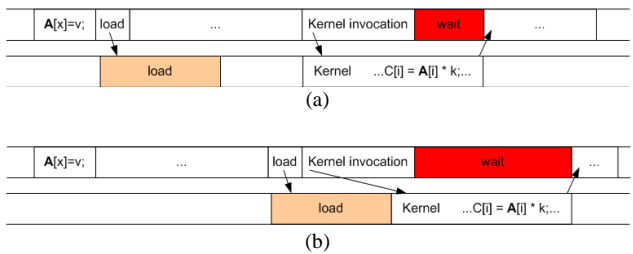

(a)

(b)

Figure 6: Advanced Load Directive Optimization. (a) Variables are loded as near as possible of the last CPU write. (b) Variables are loaded when kernel is invoked.

Moreover, OMP2HMPP can deal with context situations in which the source code contains nested loops. OMP2HMPP determines if an operation on a variable is made inside a loop and then adapts the data transfer to the proper context situation Figure 7 and Figure 9 illustrate an example with possible load the value of the variable a is required to compute the value of C. Since the last write in CPU of A is inside a loop with a different nested level than the GPU block, OMP2HMPP has to backtrack the nesting of loops to find the block shared by both loops. Then, similar than Figure 7, OMP2HMPP optimizes the load of A adding the advancedload directive as close as possible to the end of the loop. We could have the inverse problem changing the block that is computed in GPU, as shown in Figure 9. In this Figure, the result of the GPU kernel is needed in CPU to compute C that is not at the same loop level. In that case, the optimum way to add the delegatestore directive, will be just before the start of the nested loops where the computation of C is located.

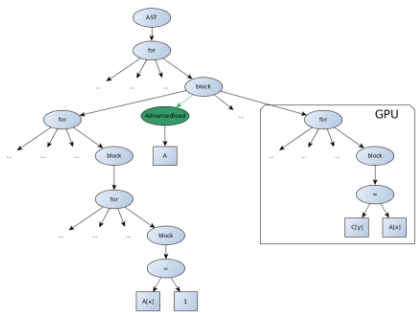

Figure 7: Data Transfer in Loops Example

Table 8: Output CSV Spreadsheet example from OMP2HMPP using Jacobi Source code.

| Version/Measure | Signature | Time Expended(ms.) | Energy Consumption(J.) |
|---|---|---|---|
| Original(OpenMP) | 0, 0, 0 | 59500 | 17428 |
| Adv_loaddelStoreNoUpdate... | 9, 1, 0 | 9611 | 3401,55 |
| Adv_loadRel... | 11, 3, 0 | 10530,2 | 3819,2 |
| Adv_loadRel... | 11, 1, 0 | 10572,4 | 4109,9 |
| Adv_loadRel... | 10, 1, 0 | 10844,4 | 3974,2 |
| ... | ... | ... | ... |

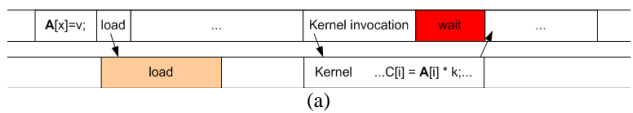

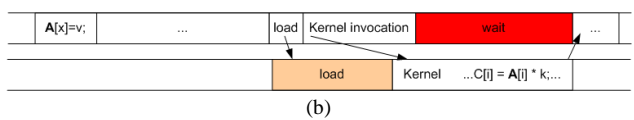

Figure 8: Delegate Store Directive Optimization. (a) Variables are download as far as possible of the kernel finish, next to the first CPU read. (b) Variables are downloaded when kernel finish.

Table 6 shows a more complex example of the loop context understanding. OMP2HMPP is able to understand more complex context inside loops situations decreasing in these unnecessary transfer repetitions between CPU→GPU and GPU→CPU. OMP2HMPP understands that a previous instructions are reading this variable. Table 5 also illustrates how OMP2HMPP downloads this variable from GPU to CPU after finishing of the current iteration in order to have the update its value, which has been modified in GPU.

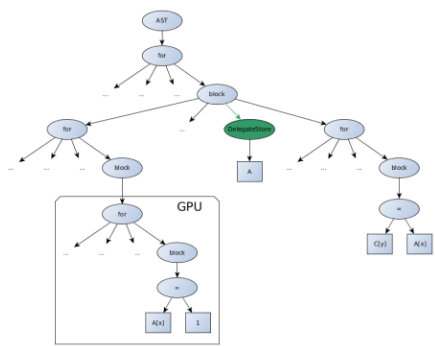

Figure 9: Data Transfer in Loops Example

We extended the problem to have a better

Table 9: Inline Transformation.

| OPENMP | HMPP |
|---|---|
| ```
void g(int &a,int b)
{
    int r=2;
    int c=1;
    a=a+r+2;
    int ret = a+b+c+r*2;
}
int f(int a)
{
    return a+1;
}

int main()
{
  int l;
  int x=2;
  l=f(1)+f(2)+g(x,6);
  l=l*g(x,2);
  return l;
}
``` | ```
int deletedFunctionBodyNamed_g = 1;
int deletedFunctionBodyNamed_f = 1;
int main()
{
    int l;
    int x = 2;
    int _p_0_f_0 = 1;
    int _return_0;
    {
        int ret_f0;
        ret_f0 = _p_0_f_0 + 1;
        _return_0 = ret_f0;
    }
    int _p_0_f_1 = 2;
    int _return_1;
    {
        int ret_f1;
        ret_f1 = _p_0_f_1 + 1;
        _return_1 = ret_f1;
    }
    int *_p_0_g_2 = &x;
    int _p_1_g_2 = 6;
    int _return_2;
    {
        int r = 2;
        int c = 1;
        *_p_0_g_2 = *_p_0_g_2 + r + 2;
        int ret = *_p_0_g_2 + _p_1_g_2 + c + r * 2;
        int ret_g2;
        ret_g2 = ret;
        _return_2 = ret_g2;
    }
    l = _return_0 + _return_1 + _return_2;
    int *_p_0_g_3 = &x;
    int _p_1_g_3 = 2;
    int _return_3;
    {
        int r = 2;
        int c = 1;
        *_p_0_g_3 = *_p_0_g_3 + r + 2;
        int ret = *_p_0_g_3 + _p_1_g_3 + c + r * 2;
        int ret_g3;
        ret_g3 = ret;
        _return_3 = ret_g3;
    }
    l = l * _return_3;
    return l;
}
``` |





understanding of the use of the contextual information of the use of group, mapbyname, noupdate or asynchronous directives. After the use of any variable by the GPU, if that variable is not written in CPU before the next read in GPU, the optimization phase will keep the variable in GPU without downloading it to CPU and automatically create a group of HMPP codelet using group directive, even if it was not specified in the original OpenMP source code. The creation of this group will represent that both kernels can share this variable by just load transfer CPU→GPU and one download transfer GPU→CPU with the use of mapbyname directive. OMP2HMPP includes option to use asynchronicity in kernel invocation when that can be beneficial. OMP2HMPP extracts information of the next usage of the kernel variables and adds the asynchronous HMPP directive taking that use in account. Finally, noupdate directive will be used to keep variables in GPU that are not updated in CPU as shown in Table 5 where we illustrate the kernel call and in Table 4 where we show the codelet transforming these variables into input parameters. This transformation keeps the variables myTable and myTableOut in GPU and just does one load/download transfer.

### 2.2 Inline Phase

This phase takes any function call in the input code and then detects his body declaration identifying the declaration of the needed function parameters, to create a block of code that has the same meaning. At the same time this phase is responsible for checking the scope of the variable to detect any wrong usage of global variables inside HMPP codelet.

We can see in Table 9 an example of this procedure. Left sideshows the original code, and the right one the code after inline transformations.

We add a declaration of all the needed parameters inline is added to change their name inside this block by the pattern name p x f y, where x is the position of the parameter in the function call, f is the name of the inlined function, and y is an index in order to avoid re-declaration. The index y increases each time that a function is inlined. In addition, we declare a new variables ret_fy, ret and _return_y , these variables deal with return parameters of the original function. The inline phase divides an expression which is formed by a mathematical expression of the results of a set of function calls, each of the function call. All results of the function calls are stored in return y variables and then the original expression is computed using these new variable values.

In the first lines of the code transformed by OMP2HMPP appear that global informative variables. These variables are created in order specify to the programmer which are the functions that have been inlined.

## 3. Results

The elapsed time and energy consumption used for the parallel execution of the code with different options are presented in the file report file generated by OMP2HMPP tool. This file is a comma separated value file that has all the information needed in order to do its further analysis. OMP2HMPP can do several executions of each of the generated versions on different input codes to extract the median of time and energy spends in their execution.

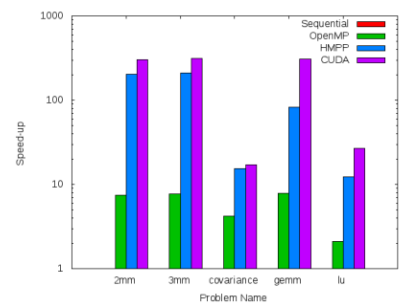

Figure 10: Speed Up Comparison

Table 8 illustrates an example of CSV results in a spreadsheet application. The first column shows the name of the generated file, the second column a unique signature (referred to the selected version of HMPP directives by the use of OMP2HMPP FIXED directive), and in the following columns the values for the time and energy measurements.

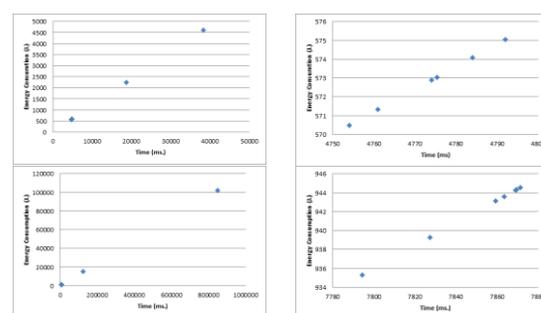

Figure 11: Energy/Time Trade-off. (a) LU. (b) LU Detailed. (c) GEMM. (d) GEMM Detailed.





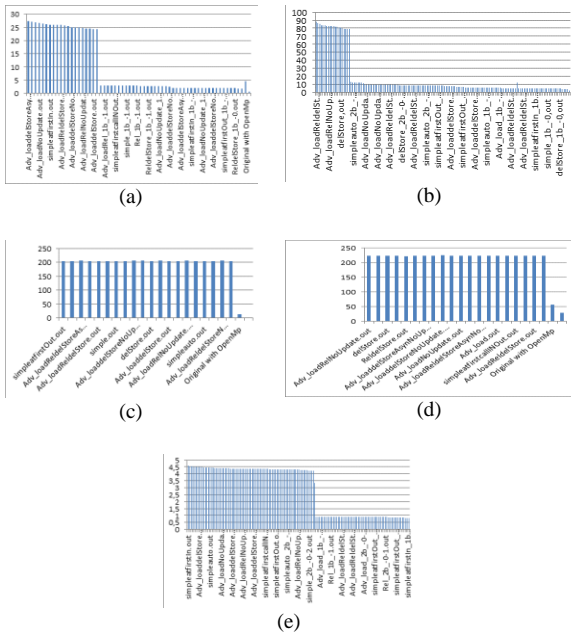

Figure 12: GOPS/W. (a) 2MM. (b) 3MM. (c) LU. (d) GEMM (e) Covariance

The generated versions were executed in B505 blade, equipped with 2 quad-core Intel Westmere-EP E5640 2.66 GHz CPUs, 24 GB of memory, and 2 Nvidia Tesla M2050 GPUs. This blade is equipped with energy meters which can be accessed by the BMC (Baseboard Management Controller), the embedded micro-controller that manages the blade: power-on/off of the blade, its temperature, the ventilation, and the energy consumption. The energy consumed by the components of the chassis (AC-DC transformer, Chassis management module, Interconnect switches, etc.) is not taken into account since they are outside the blades. The energy consumption is given in Watt-hour (Wh.) we can transform this values to Joules applying the corresponding factor conversion $3600 Joule = 1Wh$. The measured energy includes the following units:

—Active CPU
—Idle CPU
—Memory
—GPUs

An example of performance analysis of the codes generated by OMP2HMPP, is performed on a set of codes extracted from the Polybench [23] benchmark and then. We compare the execution of OMP2HMPP resulting codes with the original OpenMP version, and also with a hand-coded CUDA version and with a sequential version of the same problem. Figure 10 shows the speedup comparison for the selected problems. This figure shows that OMP2HMPP produces good transformation of the original OpenMP code, obtaining an average speed up of $113\times$. The best speedup of an automatically generated version is still a bit lower that the obtained for the CUDA hand-coded code version with that has an average speed up of $1.7\times$. Moreover, the average speedup obtained when we compare the generated code to the original OpenMP version is $31\times$ that is a large gain in performance for a programmer that additionally does not need to have any knowledge in GPGPU programming.

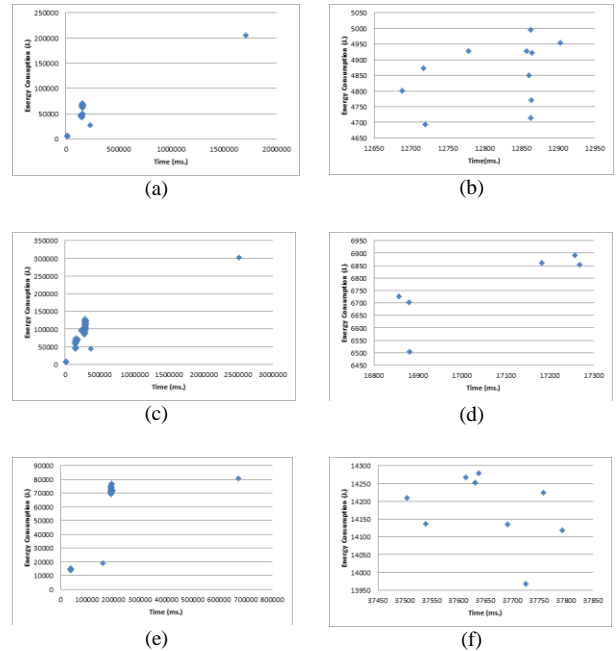

Figure 13: Energy/Time trade-off. (a) 2MM. (b) 2MM Detailed. (c) 3MM. (d) 3MM detailed. (e) Covariance. (f) Covariance Detailed.

OMP2HMPP measured energy and time for all the generated of all the problems in the benchmark. Figure 13 and Figure 11 shown those measurements and allows selecting the right implementation according to the desired working point. The left column in these figures shows the full set of generated versions and the right one detail the best ones. Figure 13 and Figure 11 illustrate that some of the cases there is a real trade-off between energy and time.

Finally, Figure 12 presents the energy efficiency (in GOPS/w) for all cases. These results manifest that the generated versions increase the number of operations per watt that can be done after the re-factorization done by OMP2HMPP.

## 3. Conclusions

We have built an OMP2HMPP source to source compiler oriented to provide GPGPU programmers a powerful tool that will facilitate the study of all the transformations that can be done of an OpenMP block into





any possible HMPP codelet and callsite. These transformations can be compared at the same time with the non-transformation of the proposed block that combines CPU and GPU by using parallel shared memory computation (CPU, GPGPUs). With our automatic transformations the programmer avoids learning the meaning of HMPP directives and, more important, obtains a good performance analysis that allows a smart selection of the best version according to its requirements. By using OMP2HMPP on the Polybench benchmark subset we obtain an average speed up of 31x and an average increase of energy efficiency of 5.86x, comparing the generated version with the best results coming from OpenMP version. OMP2HMPP tool produces solution that rarely differs from the best HMPP hand-coded version. We notice that a CUDA hand-coded version CUDA obtains a speedup near 1.7x compared with the version the best speedup of OMP2HMPP. The automatic translation provided by our tool can be also useful for experimented users that want to have, with minimal effort, a GPGPU flavor of the code partitioning and mapping for any given problem from which better performance with additional clever transformations. Current version of the tool has some limitations with the OMP2HMPP expansion. One of these limitations is that in the actual version of OMP2HMPP we generate a maximum of twenty-one possible configurations for each simple OpenMP block and this number of generated version grows exponentially when we try to transform many simple OpenMP block at the same time or a groups of OpenMP parallel blocks. This is caused because the versions that OMP2HMPP proposes are either considering the possibility of full HMPP source codes or creating possible solutions that combine the use of OpenMP and HMPP. Then, its result is a huge number of versions that can complicate considerably the final analysis. OMP2HMPP will solve this issue in future versions, to minimize the execution time needed for higher number of version. This will be done by improving the optimization stage with the capability to delete some redundant versions or a priori inefficient versions. Then, OMP2HMPP propose the smallest set of possible versions with the smarter pragma combination of OpenMP and HMPP.

OMP2HMPP will do this task separating each of the OpenMP annotated pragma and creating from those a new program with their outline their context. After that, each of this sub-programs will be compiled and executed to establish if it is efficient (in terms of its potential to obtain good results) to optimize the CPU parallel block under a pre-established metric. This capability will save testing time and OMP2HMPP still will give to the programmer the versions to check the best trade-off between execution time and energy consumption.

## 4. Acknowledgments


This work was partly supported by the European cooperative ITEA2 projects 09011 H4H and 10021 MANY, the CATRENE project CA112 HARP, the Spanish Ministerio de Economía y Competitividad project IPT-2012-0847-430000, the Spanish Ministerio de Industria, Turismo y Comercio projects and TSI-020100-2010-1036, TSI-020400-2010-120. The authors thank BULL SAS and CAPS Entreprise for their support.

**Albert Saà-Garriga** received his B.Sc. degree in Computer Science and M.Sc. degree in Computer Vision and Artificial Intelligence from Universitat Autònoma de Barcelona (UAB), Bellaterra, Spain. He is currently at CEPHIS (Hardware-Software Prototypes and Solutions Lab), research center at the UAB, where he is doing his Ph.D. studies. His main research interests include parallel computing, source to source compilers and computer vision systems.

**David Castells-Rufas** received his B.Sc. degree in Computer Science from Universitat Autònoma de Barcelona. He holds a M.Sc. in Research in Microelectronics from Universitat Autònoma de Barcelona. He is currently the head of the Embedded Systems unit at CAIAC Research Centre at Universitat Autònoma de Barcelona (UAB), where he is doing his Ph.D. studies. His primary research interests include parallel computing, Network-on-Chip Based Multiprocessor Systems, and parallel programming models. He is also associate lecturer in the Microelectronics department of the same university.

**Jordi Carrabina** leads CAIAC Research Centre at Universitat Autònoma de Barcelona (Spain), member of Catalan IT network TECNIO. He received his PhD degree from Universitat Autònoma de Barcelona. His main interests are Microelectronic Systems oriented to Embedded Platform-based Design using System Level Design Methodologies using SoC/NoC Architectures and Printed Microelectronics Technologies in the Ambient Intelligence Domain. He is a Prof. T. at Universitat Autònoma de Barcelona where is Teaching EE and CS at the Engineering School and in the MA of Micro & Nanoelectronics Engineering and Multimedia technologies, at UAB and Embedded Systems at UPV-EHU.